\documentclass[preprint]{aastex}

\usepackage{graphics}
\usepackage{natbib}
\begin{document}

\title{Observations of a Stationary Mid-Latitude Cloud System on Titan}

\shortauthors{M. \'Ad\'amkovics et al.}

\author{
  M. \'Ad\'amkovics\altaffilmark{1},
  J. W. Barnes\altaffilmark{2},
  M. Hartung\altaffilmark{3},
  and I. de Pater\altaffilmark{1}}

\altaffiltext{1}{Astronomy Department, University of California, 
Berkeley, CA, USA}

\altaffiltext{2}{Department of Physics, University of Idaho, Moscow, ID, USA.}

\altaffiltext{3}{Gemini Observatory, La Serena, Chile}


\slugcomment{Accepted for publication in {\em Icarus}, 10.1016/j.icarus.2010.03.006}

\clearpage

%
%

\noindent
\textbf{Abstract:}
We report the observation of a cloud system on Titan that remained localized
near 40$^{\circ}$S latitude and 60$^{\circ}$W longitude for at least 34 hours.
Ground-based observations obtained with the SINFONI imaging spectrograph at the 
Very Large Telescope over 4 consecutive nights recorded the lifetime and 
altitude of the unresolved cloud system. Concomitant measurements made by 
Cassini/VIMS over 3 hours resolved changes in the altitude and opacity of 
individual regions 
within the system during this time. Clouds are measured from 13 to 37\,km 
altitude with optical depths per pixel ranging from $\tau$=0.13 to 7. Short 
timescale rise times are consistent with previous measurements of the evolution
of mid-latitude clouds; however the long timescale localization of the cloud 
structure is unexplained. We speculate about the role of meso-scale circulation 
in relation to cloud formation.
\\
\\
\textbf{Keywords}: Atmospheres; Infrared Observations; IR Spectroscopy

\clearpage

%
%

\section{Introduction}

Clouds occur at various geographical locations on Titan with distinct morphologies,
which suggests that different mechanisms and meteorological conditions underly cloud formation. The diversity of clouds provides insight into Titan's climate dynamics and exotic methane-based meteorological cycle. Since the first spectroscopic indications of clouds \citep{Toon1988,Griffith1998, Griffith2000}, they have been imaged near the South pole; in the region of maximum solar insolation at the time \citep{Brown2002,Roe2002b}. The polar storm clouds seen by Cassini/ISS are clusters of individual small-scale convective clouds \citep{Porco2005}.  These southern polar clouds occasionally develop into massive storms and their frequency changes seasonally \citep{Schaller2006a, Schaller2006b, Turtle2009}. There are signs of changes on the surface after large storms, which may be the accumulation of precipitation into lakes at the south pole \citep{Turtle2009}. Clouds that are preferentially aligned near 40$^{\circ}$S latitude were first observed by \citet{Roe2005a}. These ``southern temperate'' or ``southern mid-latitude'' clouds evolve over hourly timescales \citep{Griffith2005,Porco2005}.

Clouds composed of ethane have been discovered near the North polar tropopause that likely form from Hadley cell subsidence there \citep{Griffith2006}. A thin cloud of solid methane was inferred from the relative humidity profile near the Huygens landing site, 110$^{\circ}$W, 10$^{\circ}$S \citep{Tokano2006a}, which seems to form a stratiform layer over the globe \citep{Adamkovics2007}. The characteristics of clouds in the tropics is consistent with a dry climate there now, unlike the polar atmosphere \citep{Griffith2009}. Streaky clouds have been catalogued during the Cassini mission that may be closely related to the northern polar lakes \citep{Brown2009a}. The south pole has low-altitude fog, which indicates the release of methane into the atmosphere from the surface \citep{Brown2009b}. Both seasonal changes in meteorological conditions and geographical differences between the poles could lead to the distinct cloud morphologies that are observed at various locations.

Regions of upwelling in general circulation models (GCM) of Titan \citep{Tokano2005a,Rannou2006, Mitchell2006,Mitchell2008} are consistent with the location of both southern midlatitude and southern polar clouds.  
The GCMs each predict that as the large scale motion of the atmosphere changes with seasons, so will the frequency and location of these convective clouds. There are differences among these models for when and where such changes will occur. The observed change in cloud top altitude is also consistent with convective cloud formation \citep{Griffith2006}, as is the scarcity of clouds, which coincides with the relatively small amount of convective cells compared to Earth \citep{Lorenz2003}. \citet{Barth2007} used a dynamic cloud model to reproduce the cloud-top height, horizontal extent, and short lifetimes of individual mid-latitude (or temperate) clouds observed by \citet{Griffith2006} with Cassini/VIMS. Such cloud models could be used to study the specific triggering mechanism for convection, potentially disentangling the role of synoptic-scale effects from those of global circulation. 

Measurements of the evolution of clouds on larger spatial scale and longer timescales provide observational constraints for models of cloud formation. The observations we present were made during the course of ground-based observational support of the Cassini flybys and have captured the daily evolution of a massive cloud system on Titan. We monitor the evolution of a southern mid-latitude cloud system over a broad range of timescales and suggest that meso-scale meteorology may have an impact on cloud system formation.

\section{Observations}

The ground-based observations we present were performed on 28 Dec 2006, nightly from 28 -- 31\,Jan\,2007\,UT and on 23\,Feb\,2007\,UT, with the adaptive-optics aided near-IR imaging spectrometer, SINFONI, at the Very Large Telescope (VLT). The spectrometer uses two sets of stacked mirrors to optically divide and rearrange the field of view (FOV) into a single synthetic long slit that is spectrally dispersed by a grating onto the detector. We used the $0.8'' \times 0.8''$ FOV, corresponding to a spatial pixel scale of $0.0125'' \times 0.0250$'', with the grating that covers 1.45 -- 2.45\,$\mu$m at a spectral resolution, $\Delta\lambda \sim$ 0.75\,nm (corresponding to a resolving power, $R = \lambda / \Delta\lambda \sim$ 3000). Data-reduction follows the SINFONI pipeline \citep{Modigliani2007}. Observations and reduction are detailed in  \citet{Adamkovics2007, Adamkovics2009}.

For the spacecraft observations, we use publicly available Visual and Infrared Mapping Spectrometer (VIMS)  data from the Cassini T24 flyby on 29 Jan 2007 UT. VIMS implements a spectral mapping technique \citep{Brown2004} to assemble images at 352 wavelengths, covering 0.3 to 5.1\,$\mu$m.  The data are reduced from observed data numbers to units of albedo, $I/F$, according to \citet{Barnes2007}. The VIMS observations provide high time-resolution, sampling at $\sim$30\,min intervals over the course of $\sim$3\,hrs, midway through the ground-based observations. In the first three datasets, exposure times are 480\,ms per pixel, while in the last 3 they are 440\,ms per pixel. The two complimentary sets of near-IR data are used to study the evolution of the observed cloud system on a broad range of timescales, from 30\,min to 30\,hrs (Table 1). For the ground-based data, the diffraction-limited (0.05$\arcsec$ at 2\,$\mu$m) resolution is $\sim$300km, so we make a conservative estimate of $\sim$350km at the latitude of the clouds. For VIMS, the resolution range is 100 -- 150\,km.

\section{Analysis of Spectra}

\begin{figure}
\begin{center}
\noindent\includegraphics[width=40pc]{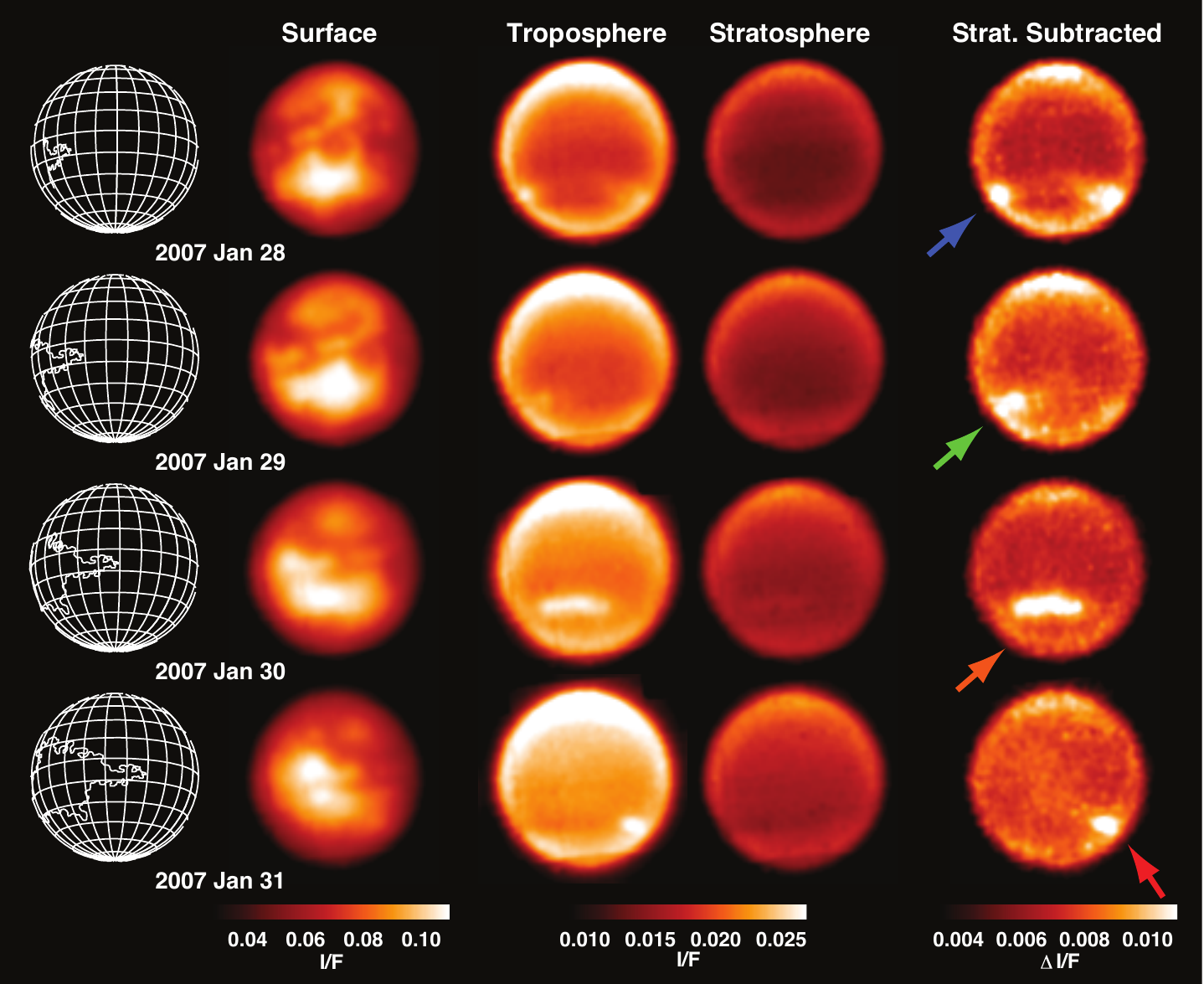}
\end{center}
\caption{\label{fobs}}
\end{figure}

The varying gas opacity at wavelengths around 2\,$\mu$m, largely due to the strength of CH$_{4}$ absorption bands, can be used to probe scattering from different altitudes in the atmosphere \citep{Griffith1991}. Images recorded through narrow bandpass filters at these wavelengths have been used to identify clouds in the troposphere, e.g., \citet{Roe2002b}. 
We extract images from the VLT/SINFONI datacubes in the following wavelength regions to find the spatial locations of clouds: the surface (2.027--2.037\,$\mu$m), troposphere (2.123--2.128\,$\mu$m), and stratosphere (2.150--2.160\,$\mu$m), as shown in Figure \ref{fobs}. Flux from stratospheric aerosol scattering is limb brightened and can obscure faint clouds near the edge of the disk. To make an unambiguous identification, we subtract the contribution of stratospheric flux from images probing the troposphere \citep{Adamkovics2007}. After subtraction, the contrast due to tropospheric scattering is increased as the limb brightening is suppressed. These images are used to identify the locations of cloud spectra in the datacubes.

\begin{figure}
\begin{center}
\noindent\includegraphics[width=30pc]{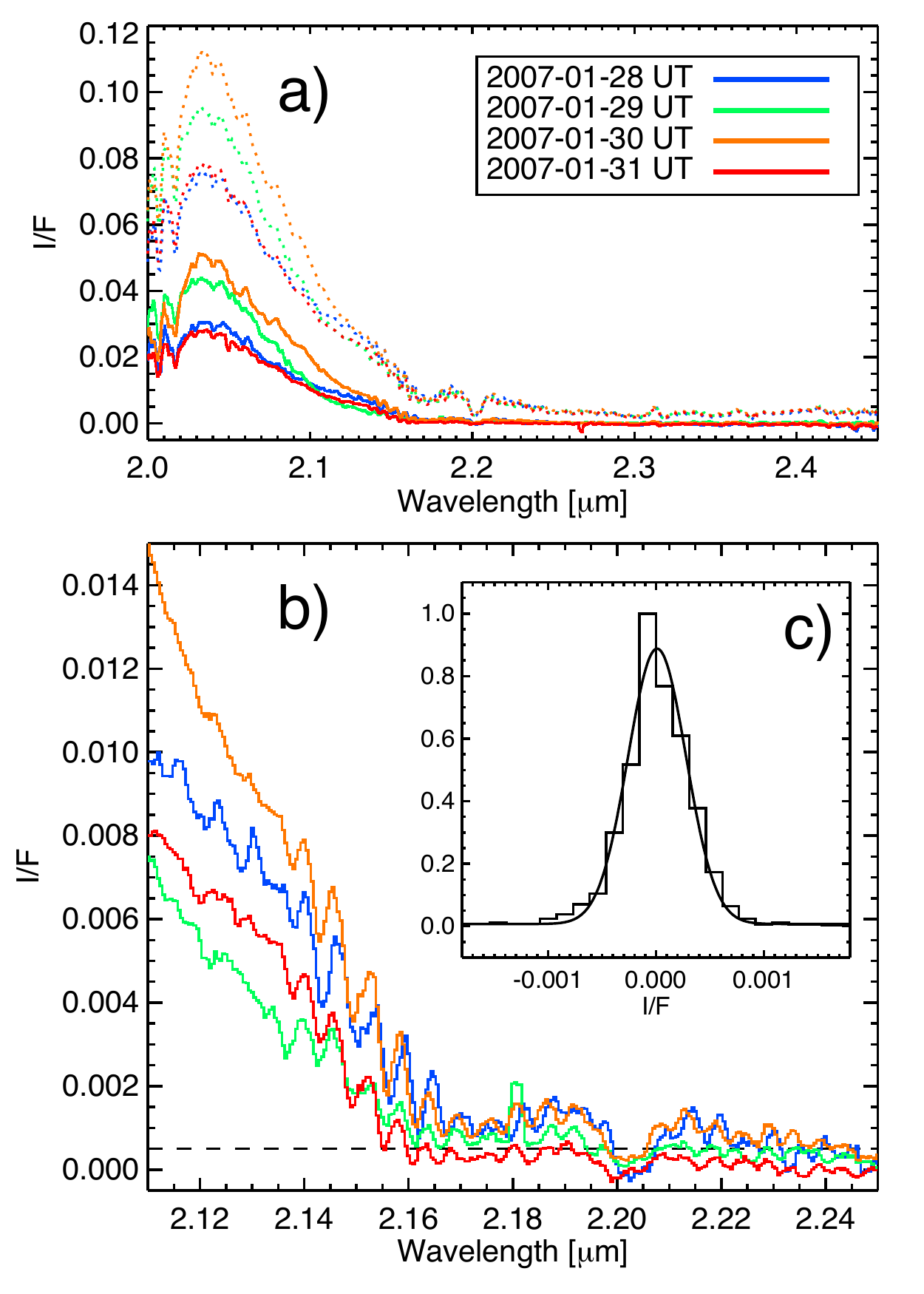}
\end{center}
\caption{ \label{fspec}}
\end{figure}

Spectra from 5$\times$5 pixel spatial regions centered on each cloud are then extracted from the datacubes and box-car smoothed over 5 spectral pixels. An initial estimate for the relative altitude of the cloud tops is determined by comparing the $I/F$ in the spectra near 2.13\,$\mu$m (Figure \ref{fspec}a, dotted lines); spatial variations in the surface reflectivity will affect the spectrum below 2.13\,$\mu$m  (Figure 2 in \citet{Adamkovics2009}) and increasing the path-length through the stratosphere will increase the observed flux at wavelengths longer than 2.15\,$\mu$m (e.g., images of the stratosphere in Figure \ref{fobs}). Nonetheless, on 28~Jan and 30~Jan~2007~UT the $I/F$ at troposphere-probing wavelengths is greater than on 29~Jan and 31 Jan 2007 UT.

We confirm that these four spectra are significantly different from one another by removing the spatial variation in stratospheric albedo using a cloud-free spectrum taken at the same location on disk, recorded on 28 Dec 2006 UT. This assumes that the atmospheric aerosol is constant within the observational uncertainty over the 1 month between observations. Clear-sky subtracted spectra are plotted in Figure \ref{fspec}a and \ref{fspec}b (solid lines), and are indistinguishable beyond 2.25\,$\mu$m, where the stratospheric scattering has been removed. The histogram of $I/F$ values over 2.30 -- 2.45\,$\mu$m, for all the corrected spectra (Figure \ref{fspec}c) is centered around zero, indicating that the stratospheric flux from the clear-sky spectra are a good proxy for the stratospheric flux in the cloudy spectra. The spread in this histogram is an estimate of the observational uncertainty in the spectrum, $\sigma_{I/F} \sim$ 0.0005, plotted as a horizontal dashed line in Figure \ref{fspec}b. At wavelengths where the $I/F$ in the corrected spectra exceeds the noise level, there is a variation in upper tropospheric albedo due to clouds. Comparison of the spectra beyond 2.16\,$\mu$m indicates that on 28 and 30 Jan 2007 UT, the tops of the clouds extended significantly higher in the atmosphere than on 29 Jan 2007 or 31 Jan 2007 UT. The lowest cloud-top altitude was observed on 31 Jan 2007 UT, contrary to the interpretation from comparing the albedo at 2.13\,$\mu$m. The wavelength region used to identify cloud locations in extracted images is not sufficient alone for altitude determination, even if a narrow spectral bandpass is used. Increased flux at 2.13\,$\mu$m does not necessarily indicated a higher cloud-top altitude. In the following section we use numerical models to support and quantify the conclusions above. The subtraction of stratospheric flux from images and the correction with clear-sky spectra are not considered when directly comparing models and observations.

\section{Cloud Altitude and Optical Depth Retrieval}
 \subsection{Clouds in radiative transfer calculations}
The characteristics of clouds are determined by simulating the observed datacubes with a radiative transfer model in which the cloud altitude and opacity are free parameters. The forward model is described in detail in \citet{Adamkovics2009}, which builds on previous studies by \cite{Griffith1991, Griffith2005}. Recently, \citet{Griffith2009} used the characteristics of tropical clouds to distinguish the warm, dry, tropical atmosphere from the relatively cooler, more humid, polar regions. A difference between these radiative transfer models is the treatment of cloud layers.

In general, an individual model layer is characterized by either the scattering from large cloud particles or from haze aerosols, but not both. Unlike the opacity due to absorption, which can been summed from various atmospheric constituents, there is no obvious method to calculate a combined scattering albedo or a composite phase function for a heterogeneously scattering layer. A common solution to this problem is to add additional layers to the model, which account for various types of scattering.

One method for adding clouds, without significantly changing the haze, is to divide the model atmosphere into enough layers that the aerosol opacity is small in each layer; replacing the haze scattering with cloud scattering, and thereby removing the scattering by haze at that altitude, should have a negligible effect in this case. The haze vertical profile is well constrained for tropical regions by the Huygens probe data \citep{Tomasko2005}, and parameterizations exist to approximate the haze profile at other latitudes \citep{Adamkovics2006}, so the haze structure is considered to be a fixed quantity in the model. Unlike previous studies of cloud top altitudes, where replacing the haze scattering in a layer by cloud particles could be compensated for by increasing the haze density somewhere else in the atmosphere, we maintain the haze vertical structure when adding two  additional layers at the altitude where clouds occur. This maintains the aerosol profile exactly and is computationally faster than including several additional layers.

In our model, the atmospheric layer that encompasses the cloud is separated into two layers, splitting the scattering and opacity equally. The cloud scattering layer is then inserted between these two layers with a single scattering albedo, $\omega$=0.99, and phase function with a Henyey-Greenstein asymmetry parameter $g$=0.85, that is distinct from the haze layers. As with previous studies \citep{Griffith2005,Griffith2009}, the clouds are assumed to be uniformly reflecting particles (constant single-scattering albedo with wavelength) that are much larger than the 2\,$\mu$m wavelength of the observations. The methodology is tested in the limit of small cloud optical depths, which  reproduce the cloud-free, or clear-sky, calculations.

\citet{Griffith2009} use a 32 degree polynomial expansion to reproduce the aerosol scattering phase function that was measured by \citet{Tomasko2008c} using the Huygens probe data at 1.5\,$\mu$m. While there are models to extrapolate the Huygens phase functions to longer wavelengths, we use a Henyey-Greenstein phase function, and likely underestimate the strongly forward-scattering peak in the phase function. Nonetheless, this approximation can reproduce both our observations \citep{Adamkovics2009} and the heat balance in the atmosphere \citep{Tomasko2008b}. It is unclear if using a closer approximation of the scattering phase function by \citet{Griffith2009} at 1.5~$\mu$m for cloud altitude discrimination outweighs the benefit of the shallower absorption band wing and lower aerosol opacity at 2\,$\mu$m. The uncertainty in the retrieval of cloud altitude, due to the degeneracy between cloud optical depth and cloud altitude discussed below, may surpass the uncertainties caused by choice of aerosol scattering phase function or bandpass. 

\subsection{Uncertainty analysis}
The approximated Henyey-Greenstein phase function allows for the rapid calculation of many spectra, with  clouds at various altitudes, spatial extents, and optical depths, so that quantitative comparisons can be made between the calculated and observed datacubes, yielding estimates of the uncertainties in the input parameters. The observational uncertainty for the entire datacube is determined by using a cloud-free model to calculate the residual at each spatial and spectral pixel in the cube. A histogram of the residuals from the entire dataset is used to approximate the single pixel uncertainty as 
$\sigma$=0.0045. This estimate of $\sigma$ is consistent with the uncertainty in the spectra presented in Figure \ref{fspec}, which were averaged spatially and spectrally. The uncertainty, $\sigma$ is used to determine a quality factor, $\widehat{\chi^{2}}$, for the model according to 
$$\widehat{\chi^{2}} = \frac{1}{N} \sum_{i}^{N} \frac{r_i^{2}}{\sigma^{2}} $$
where the indices, $i$, cover the cube with $N$ datapoints. The residual, $r_{i}$, is the difference between the calculation and the observations, $r_{i} = (I/F)_{obs} - (I/F)_{calc}$ at each datapoint. Models are calculated for a range of input parameters, resulting in a range of  
$\widehat{\chi^{2}}$ values, the smallest of which, $\widehat{\chi^{2}}_{min}$, generally corresponds to the input parameters that best reproduce the observations. In order to measure the uncertainty of the input parameters, we report values of 
$\Delta\widehat{\chi^{2}} = \widehat{\chi^{2}} - \widehat{\chi^{2}}_{min}$,
which can be plotted to illustrate the uncertainty in the model, e.g, Figure \ref{fCloudFit}. The 1$\sigma$ uncertainties in input parameters correspond to the $\Delta\widehat{\chi^{2}}$ = 1 contours.

\subsection{Cloud characteristics}

\begin{figure}
\begin{center}
\noindent\includegraphics[width=40pc]{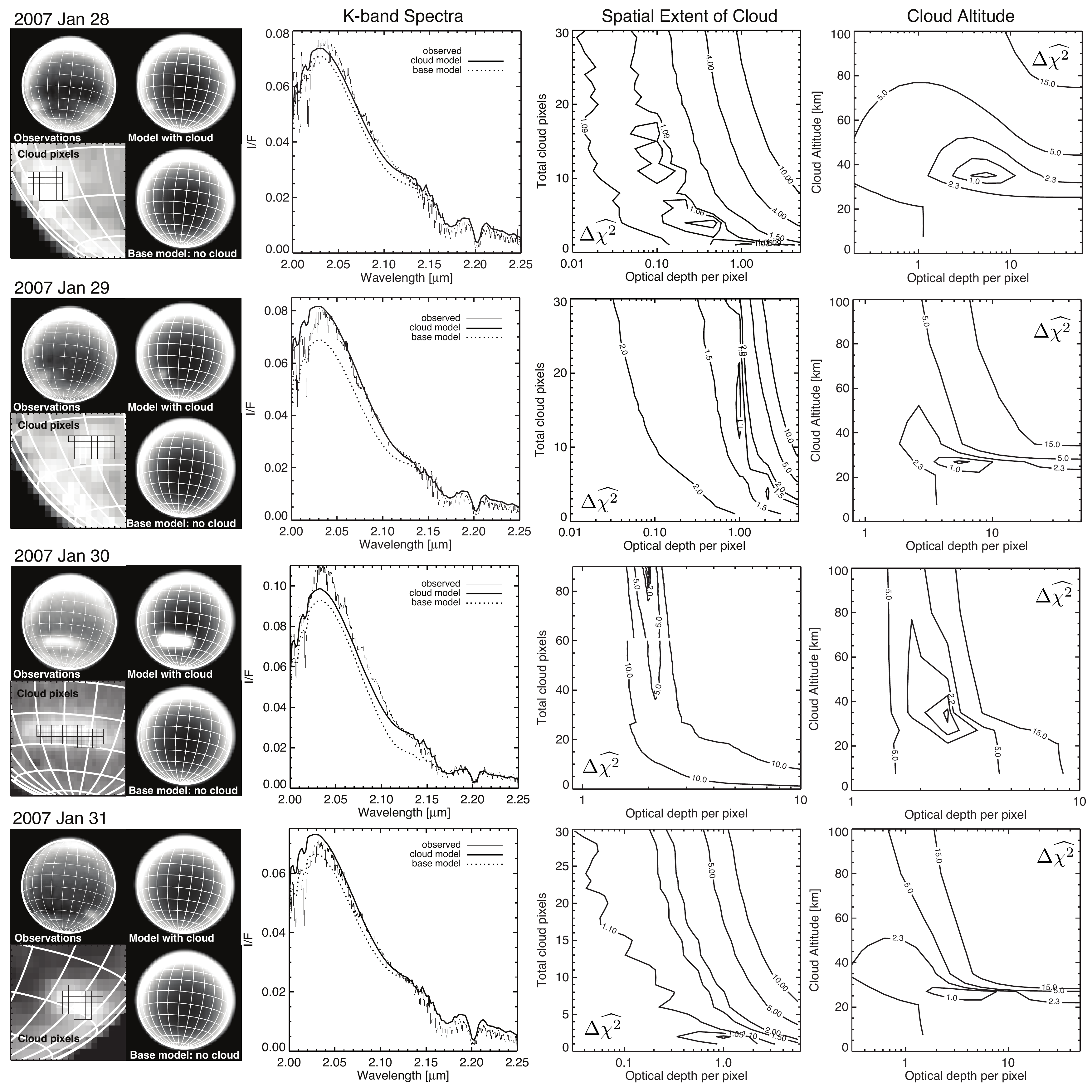}
\end{center}
\caption{\label{fCloudFit}}
\end{figure}

There is a degeneracy in the model between the spatial extent and optical depth of an unresolved cloud. Observations are compared with models that are convolved with an approximation of the point-spread-function \citep{Adamkovics2009}. A 2-dimensional Gaussian is fit to the point spread function (PSF) of a telluric standard star and is used to convolve the models. Therefore a model with a single pixel that has a large cloud optical depth is convolved to the same result as a model with a few cloud pixels of smaller optical depth. We quantitatively test how sensitive we are to this degeneracy by using up to 30 cloud pixels for each cloud, and up to 90 pixels on the night of 30 Jan 2007 UT. Results are presented in Figure \ref{fCloudFit}, illustrating that the cumulative cloud optical depth is generally maintained, regardless of the total number of pixels used to model the cloud. This is consistent with an unresolved cloud, except on the night of 30~Jan~2007 UT. Images probing the troposphere are extracted from calculations of both cloudy and cloud-free models and presented with observations in Figure \ref{fCloudFit}.
 
After evaluating the spatial extent of the cloud, we use a single cloud pixel (except on 30 Jan 2007 UT, when the cloud is extended) to model and retrieve the altitude and optical depth of the cloud. The best fitting spectra from the center of the cloud are presented with clear-sky models and the observations in Figure \ref{fCloudFit}, showing the improved fit with a tropospheric cloud at a given altitude.

On 29~Jan~2007~UT, the cloud system was observed by VIMS on an outbound trajectory. Six datasets were acquired over the course of approximately 3 hours. Images at 2.12\,$\mu$m are shown at the top of Figure \ref{fVIMSobs}. These observations are reprojected onto the same observing geometry, stratosphere-subtracted to enhance contrast, and plotted on a rainbow color scale to highlight the changes in cloud albedo. Spectra from three locations on clouds and a clear-sky position are presented at the bottom of Figure \ref{fVIMSobs} in panels A -- D. 

These spectra from the VIMS spacecraft are interpreted with techniques analogous to the ones used for the ground-based spectra, however, a discrete ordinates code, DISORT version 2.0 \citep{Stamnes1988}, is used with 8 streams to solve the radiative transfer due to the relatively high phase of the observations. The spectral range is extended to 3.5\,$\mu$m, covering regions that are inaccessible from the ground.
Wavelengths up to 2.8\,$\mu$m are well-modelled using CH$_{4}$ absorption coefficients from either \citet{Irwin2005} or using the HITRAN2008 database for line-by-line calculations of the opacity. The region beyond 2.8\,$\mu$m in the calculated spectra agrees with the observations in the clear-sky base model, but disagrees after the inclusion of clouds. Neither source of CH$_{4}$ opacity reproduces the observations at these wavelengths. This may be due to an underestimate of the gas opacity above the cloud layers, or a decrease in cloud reflectivity, at these wavelengths. 

Results for the retrieval of the cloud altitude and optical depth at three locations are presented in Figure \ref{fVIMSchi}. The best fit models for these 18 spectra, along with corresponding clear-sky models are shown in Figure \ref{fVIMSspecfit}. We approximate the per pixel uncertainty of the VIMS observations to be $\sigma$=0.003~$I/F$, however, if this is over-estimated, then the uncertainty in the input parameters will be under-estimated. A tabulated list of cloud characteristics is presented in Table \ref{tbl2}. There is a cloud on the Eastern limb on 28~Jan~2007~UT, located near 301$^{\circ}$W, 39.5$^{\circ}$S. It is unclear how, if at all, this cloud is related to the evolution of the cloud system we discuss below. 

On the four nights of ground-based observations, 28-Jan through 31-Jan, the cloud brightness increases the disk-integrated albedo at 2.12\,$\mu$m by 0.2\%, 0.6\%, 4.1\% and 1.6\%, respectively --- determined using the best-fit and clear-sky models of the observed datacubes. Brightening of less than $\sim$1\% at 2.12\,$\mu$m is consistent with the daily clouds that were observed spectroscopically by \citet{Griffith2000}, and typical of south polar clouds (e.g,. \citet{Brown2002,Bouchez2005}). Typical southern temperate clouds \citep{Roe2005a,Roe2005b} brighten  Titan by 1\% or less at 2.12\,$\mu$m \citep{Schaller2009}. The 4.1\% increase in flux observed on 30-Jan is particularly bright for southern temperate clouds, although large tropical cloud outbursts can brighten Titan by 9\% \citep{Schaller2009}. Storms near the south pole cause a fractional increase in flux of 2 -- 18\% \citep{Schaller2006a}, up to a record of 60\% \citep{Griffith1998}.

A cloud system appears to be localized near 60$^{\circ}$W through the first two nights of observations. It is unresolved during this time and occurs near the limb on the first night. Taking into account the uncertainty in longitude (Table 2), the observations are consistent with the cloud system moving less than 7$^{\circ}$.  Near 39$^{\circ}$S, this corresponds to a distance of 245\,km on the surface. Over a 24 hour period, the maximum cloud system velocity is 3\,m/s. Although it is unclear if the tropospheric winds measured by Huygens \citep{Bird2005} apply at southern temperate latitudes, the cloud system velocity is well below the zonal wind speeds ($>$10\,m/s) measured by the Doppler Wind Experiment at altitudes above 30\,km, where the cloud tops are observed on the first night. The cloud system velocity is consistent with the wind speeds measured near the surface. 

On the third night of observing, there is a resolved zonal elongation of the cloud system. While the zonal wind speeds measured near 25\,km altitude are sufficient (~$\sim$5\,m/s) to elongate clouds over 400\,km in the course of 24\,hours, the winds would have to be blowing in both directions simultaneously. The elongated region of cloud formation is centered near 60$^{\circ}$W (where the cloud system was centered the two night before), yet extends 15$^{\circ}$ to both the East and the West. If the zonal winds were the source of the large aspect ratio for the cloud system on 30 Jan 2007 UT, then it is unclear why the cloud system was not spread by winds the night before. Presumably the upper tropospheric winds would be similar from one day to the next. The short-term changes in clouds measured by VIMS (Table 2), with decreasing cloud-top altitude and increasing opacity with time, further suggest that the zonal winds may not be the sole source of the observed elongation of the cloud system during this time.

\section{Synoptic and Mesoscale Meteorology}

Each of the images with clouds is mapped onto an orthographic projection in Figure \ref{frepro}. The meridional localization and dramatic zonal elongation of the cloud system (Figure \ref{frepro}D) may be diagnostic of the synoptic scale meteorology at these latitudes. On Earth, cloud structures of this size occur in mesoscale convective systems (MCS) \citep{Houze2004}. A limitation in comparing the well-known meteorology of Earth to remote observations is the PSF, which can result in similar images when viewing bright, closely-spaced yet separated point sources (lines of small clouds) on one hand, or stratiform cloud structures on the other. A further uncertainty is that we do not have a measurement of either the convective or stratiform precipitation below the clouds that is characteristic of MCS. Nonetheless the lifetime, evolution, and extent of this mid-latitude cloud system suggest that a comparison to an MCS is relevant.

One interpretation of the linear morphology of multiple convective cells is the dynamics of a squall line. We likely observe a zonally elongated region of cloud formation, rather than an individual cloud top that is spread by winds. 
If this analogy holds, then the line of clouds would be formed by warm tropical air moving over colder air, which is predicted by GCM \citep{Rannou2006}. The tropical air is expected to be dry, since there may not be enough energy reaching the surface to evaporate methane \citep{Griffith2008}, so simple upwelling of tropical air would not necessarily form a system of clouds. Rainfall in the tropics may also be rare, and if so, the soil could be too dry to be a source of methane \citep{Mitchell2008}. For large initial methane reservoirs at the surface, \citet{Mitchell2008} models a buildup of soil moisture at 40$^{\circ}$S. The entrainment of higher-latitude air could also supply the methane and the energy required to form an MCS, if circulation can supply moist air to the mid-latitudes. Large storms have been observed near the South Pole \citep{Schaller2006a}, lakes have been observed to change \citep{Turtle2009}, and there is currently sufficient insolation to evaporate moisture from those lakes \citep{Griffith2008}. 

The length scale of mid-latitude MCSs on Earth are limited by the Rossby radius. On Titan however, this length scale is significantly larger due to Titan's slow rotation, so the thermodynamic characteristics of the boundary layer --- e.g., the total convective available potential energy (CAPE) and near-surface humidity --- will set the cloud system size and lifetime \citep{Houze2004}. A large value for CAPE may be indicative of a boundary layer that is warm and moist throughout its depth. On Earth, some of the largest MCSs form over the oceans where there is abundant moisture \citep{Houze2004} and diurnal cooling is smaller than over land, such that convection can continue overnight. On Titan, the large thermal inertia and long radiative response time in the troposphere indicate that diurnal cooling should not be significant and convection likely occurs overnight.

In the coupled GCM and cloud microphysics model of \citet{Rannou2006} the lifetime of individual large clouds is limited by the sedimentation time scale (several terrestrial days), while the lifetime of small clouds is determined by the evaporation time scale (hours). The sedimentation time scale is the scale height, $H$, divided by the sedimentation speed. A zonal cloud elongation of a few hundred kilometers (2$\times 10^5$\,m) is roughly consistent with a lifetime of $10^5$\,s and the predicted zonal windspeed within the cloud, (2.5m/s) \citep{Rannou2006}. However, the individual cells are shorter lived as indicated by the cloud-top altitudes, and multiple cells suggest that an elongated region of cloud formation, perhaps due to synoptic scale motions and not the spreading of a single cloud feature, is the cause of the zonal elongation of the cloud system observed here.

Mesoscale waves, which include gravity waves and inertial oscillations, have a variety of generation mechanisms and can organize banded regions of convective activity. While the release of large amounts of latent heat can supply the energy to organize convection, mesoscale convective regions can occur on Earth when a strong forcing is absent. For example, gravity waves in the troposphere may be confined by overlying unstable layers leading to a banded zone of convection \citep{Lindzen1976}. 

\section{Discussion}

After the first observations of mid-latitude clouds, \citet{Roe2005a} described the possible link between circulation and the latitude of cloud formation, noting that the seasonal change in circulation should lead to changes in the locations of clouds. Dynamical models of circulation were used to support this hypothesis \citep{Rannou2006} and to quantify predictions of the seasonal changes in convection and precipitation \citep{Mitchell2006}. In Jan 2007, more than four years after southern summer solstice (SSS) on Titan, the mean daily isolation is greatest near 25$^{\circ}$S, and yet the formation of clouds is still observed at essentially the same latitude as when the south polar clouds were first imaged. While this is not inconsistent with the range of latitudes where convection is predicted to occur \citep{Mitchell2006}, the preferred formation of clouds near 40$^{\circ}$ \citep{Roe2005b,Turtle2009} and the modest seasonal change in this preferred latitude, is smaller than expected \citep{Brown2009c,Rodriguez2009}.

If the border of the tropical vs. polar climate regions is $\sim$60$^{\circ}$ \citep{Griffith2008}, then clouds at $\sim$40$^{\circ}$S should be in a region that is well-approximated by the relative humidity profile measured by Huygens \citep{Niemann2005, Fulchignoni2005}. But if the distribution of moisture is not meridionally uniform --- perhaps because of the availability of energy to evaporate methane \citep{Griffith2008} or because of the seasonal cycles of precipitation and a large reservoir of methane in the soil \citep{Mitchell2008} --- then methane availability may be an additional factor that confines cloud formation to a particular latitude band. Entrainment of methane from moist polar air, or sources of moisture near the surface \citep{Roe2005b}, would both result in localized enhancements of methane. While there is no overall longitudinal preference for the mid-latitude clouds \citep{Brown2009c,Turtle2009}, the localization of cloud formation may suggest a non-negligible interaction between the surface and the atmosphere at this time.

\begin{figure}
\begin{center}
\noindent\includegraphics[width=30pc]{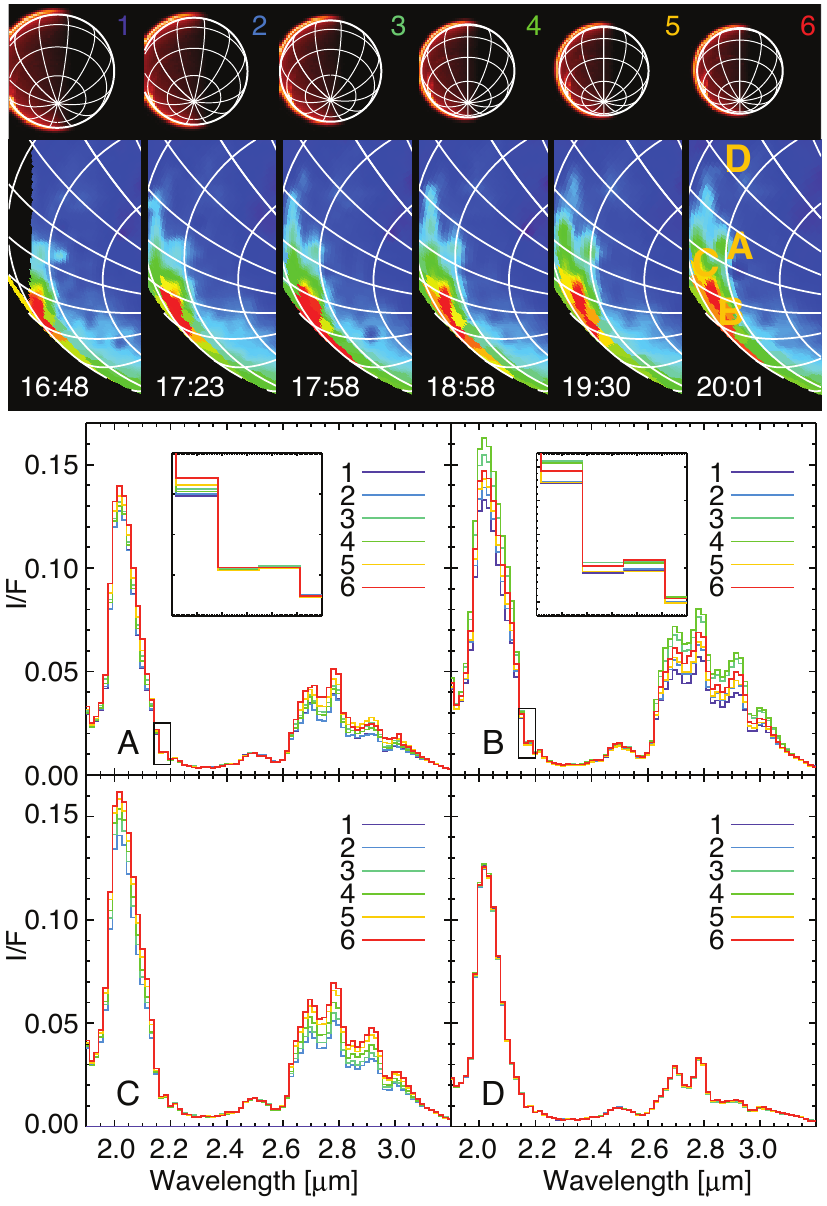}
\end{center}
\caption{\label{fVIMSobs}}
\end{figure}

\begin{figure}
\begin{center}
\noindent\includegraphics[width=25pc]{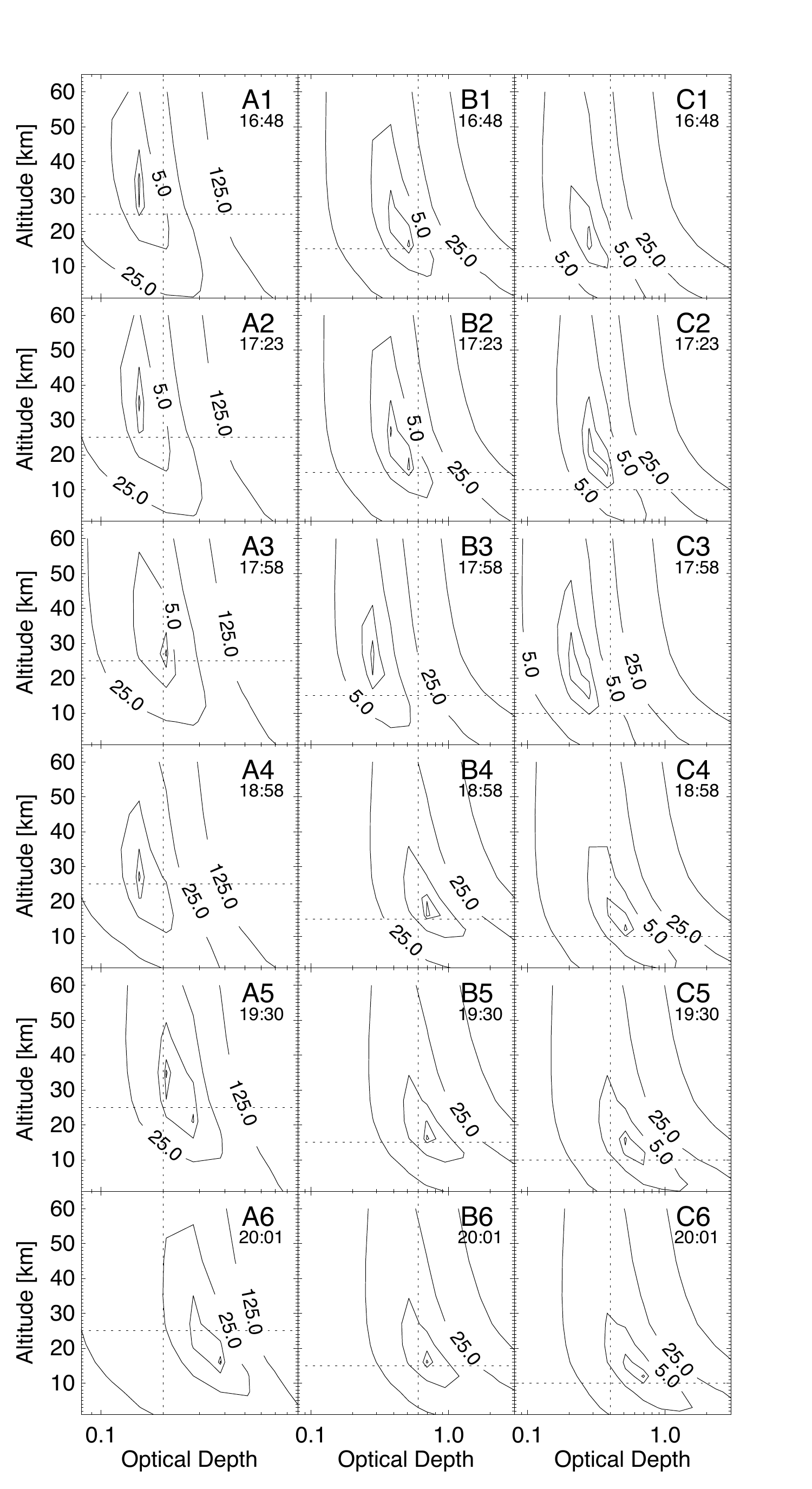}
\end{center}
\caption{\label{fVIMSchi}}
\end{figure}

\begin{figure}
\begin{center}
\noindent\includegraphics[width=40pc]{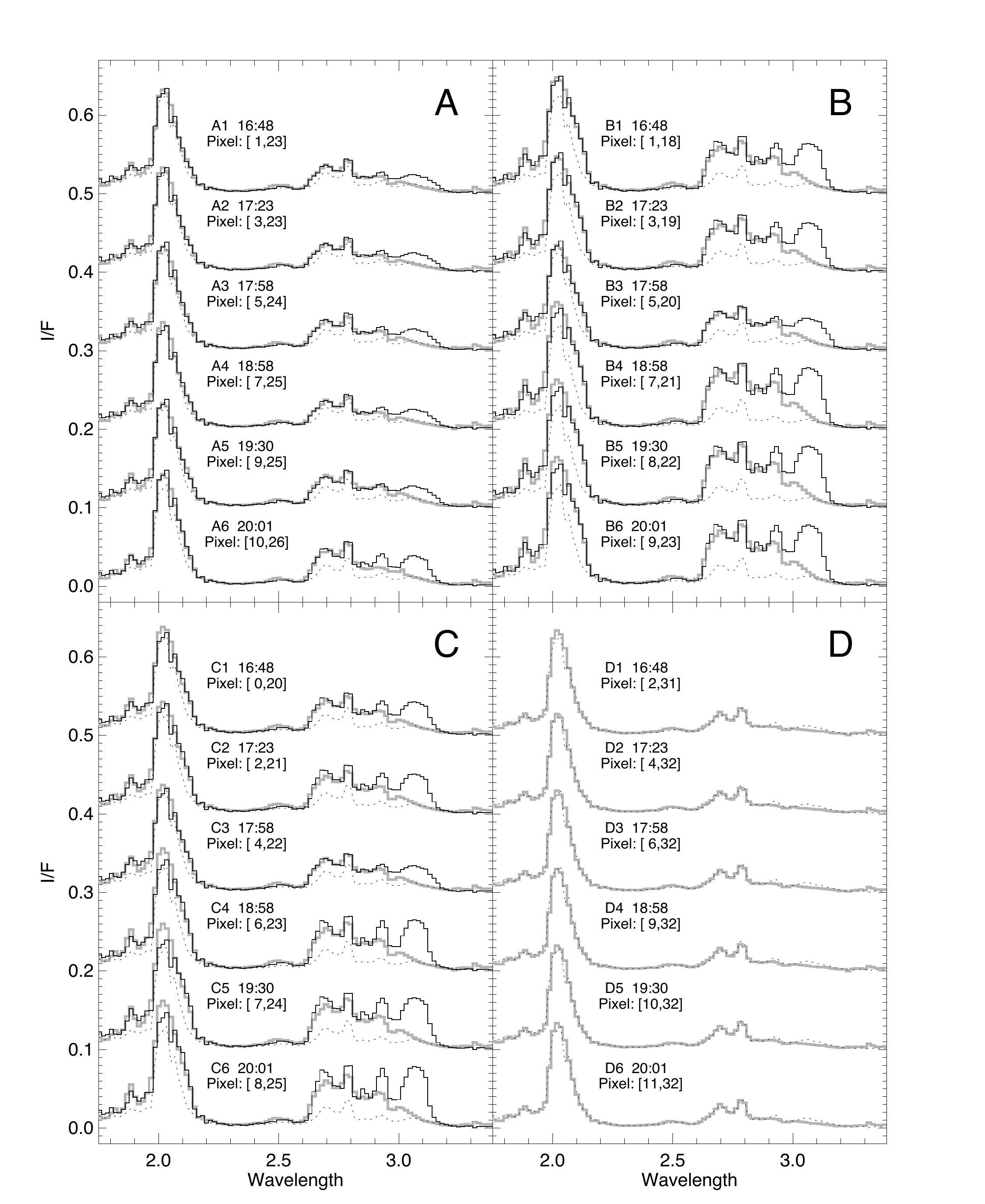}
\end{center}
\caption{\label{fVIMSspecfit}}
\end{figure}

\begin{figure}
\begin{center}
\noindent\includegraphics[width=22pc]{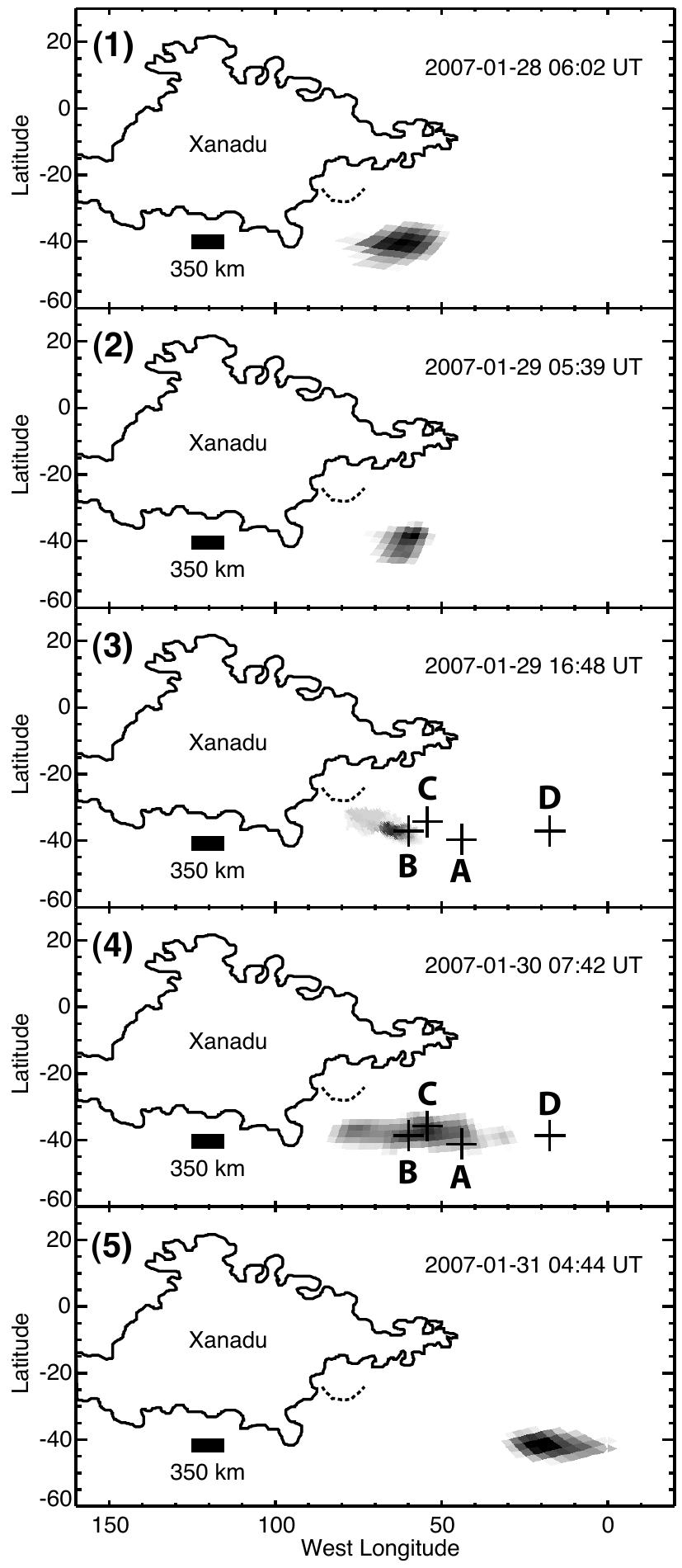}
\end{center}
\caption{\label{frepro}}
\end{figure}

\citet{Barth2007} describe how nearby cells compete for and quickly exhaust CAPE within one convective cycle. This contrasts our observations (Figure \ref{fVIMSobs}), where multiple cells are observed near each other directly prior to a large storm outburst on 29 Jan 2007 (Figure \ref{fobs}). If large scale circulation is resupplying the necessary CAPE \citep{Barth2007}, then the evolution of this cloud system may be constrained by the amount of energy needed, and therefore the mid-latitude circulation. \citet{Schaller2009} hypothesize an atmospheric teleconnection of tropical and south polar cloud formation, occurring over roughly two weeks, mediated by Rossby waves. In our observations, a region of southern temperate cloud formation persisted for at least four consecutive days, remaining stationary for three days, and did not correspond with cloud formation near the south pole. Tropical clouds may require a fundamentally different formation mechanism than temperate clouds.

We have presented observations of the spatial localization of a southern mid-latitude cloud system on Titan, with retrievals of the cloud-top altitude and opacity using both ground-based (VLT/SINFONI) and Cassini/VIMS data. The unresolved, possibly stratiform, cloud system is composed of multiple cloud forming cells that are likely driven by convection, with moist air that may be supplied from higher latitude regions, where insolation is sufficient to cause evaporation from the surface. The mechanisms for southern mid-latitude cloud formation may be related to the mesoscale dynamics of large convective systems on Earth.

\bibliographystyle{apalike}
\bibliography{refs}

\clearpage

\begin{table*}
\begin{center}
    \small

	\begin{tabular}{lcccccc}
	\hline
	\hline
		Obs. Date   & Start & \multicolumn{2}{c}{Sub-observer Point} & \multicolumn{2}{c}{Sub-solar Point} \\ 
		  (UT)      & Time  & $^{\circ}$W Long. & Lat. & $^{\circ}$W Long. & Lat.    \\ \hline
        VLT/SINFONI &       &        &        &        &         \\
		2006 Dec 28 & 08:42 & 22.5   & -12.3  & 26.8   & -14.1   \\ 		
		2007 Jan 28 & 06:02 & 1.6    & -13.2  & 3.1    & -13.7   \\ 
		2007 Jan 29 & 05:39 & 23.9   & -13.2  & 25.3   & -13.7   \\ 
		2007 Jan 30 & 07:42 & 48.5   & -13.2  & 49.8   & -13.7   \\ 
		2007 Jan 31 & 04:44 & 68.3   & -13.3  & 69.5   & -13.7   \\	\hline  
        Cassini T24 &       &        &        &        &         \\
        2007 Jan 29 & 16:48 & 311.0  & -55.29 & 36.88  & -13.7   \\
        2007 Jan 29 & 17:23 & 311.5  & -55.35 & 37.42  & -13.7   \\
        2007 Jan 29 & 17:58 & 312.1  & -55.41 & 37.97  & -13.7   \\
        2007 Jan 29 & 18:58 & 313.0  & -55.49 & 38.88  & -13.7   \\
        2007 Jan 29 & 19:30 & 313.5  & -55.52 & 39.38  & -13.7   \\
        2007 Jan 29 & 20:01 & 314.0  & -55.55 & 39.87  & -13.7   \\

		\hline  

	\end{tabular}
	
\caption{Observation dates and viewing geometry.\label{tbl1}}

\end{center}

\end{table*}

\clearpage

\begin{table*}
\begin{center}
    \small
	\begin{tabular}{lcccc}
	\hline
	\hline
    Time   & \multicolumn{2}{c}{Location} & Altitude & Optical     \\ 
    (UT)   & $^{\circ}$W Long. & Lat.   & (km)      &  Depth       \\ \hline
  	28-Jan 06:02 & 55 -- 65   & -39.2   & 37 $\pm$ 5  &  6 $\pm$ 4     \\ 
  	29-Jan 05:39 & 58 -- 62   & -38.5   & 26 $\pm$ 4  &  7 $\pm$ 3     \\ 
    30-Jan 07:42 & 45 -- 75   & -39.5   & 34 $\pm$ 9  &  2.5 $\pm$ 0.5 \\ 
    31-Jan 04:44 & 22 -- 25   & -41.2   & 28 $\pm$ 3  &  5 $\pm$ 3     \\	\hline  
    Cassini T24 - A  & 42.4 & -42.9 &   &             \\ \hline  
    29-Jan 16:48 &  &  & 34 $\pm$ 9   & 0.13 $\pm$ 0.01   \\
    29-Jan 17:23 &  &  & 36 $\pm$ 10  & 0.13 $\pm$ 0.01   \\
    29-Jan 17:58 &  &  & 29 $\pm$ 4   & 0.20 $\pm$ 0.01   \\
    29-Jan 18:58 &  &  & 28 $\pm$ 7   & 0.13 $\pm$ 0.01   \\
    29-Jan 19:30 &  &  & 32 $\pm$ 6   & 0.21 $\pm$ 0.01   \\
    29-Jan 20:21 &  &  & 20 $\pm$ 4   & 0.35 $\pm$ 0.03   \\ \hline  
    Cassini T24 - B  & 60.6 & -38.7 &        &        \\ \hline  
    29-Jan 16:48 &  &  & 21 $\pm$ 4   & 0.45 $\pm$ 0.05   \\
    29-Jan 17:23 &  &  & 32 $\pm$ 3   & 0.45 $\pm$ 0.05   \\
    29-Jan 17:58 &  &  & 25 $\pm$ 4   & 0.30 $\pm$ 0.05   \\
    29-Jan 18:58 &  &  & 18 $\pm$ 2   & 0.75 $\pm$ 0.10   \\
    29-Jan 19:30 &  &  & 18 $\pm$ 2   & 0.75 $\pm$ 0.05   \\
    29-Jan 20:21 &  &  & 17 $\pm$ 2   & 0.75 $\pm$ 0.05   \\ \hline 
    Cassini T24 - C  & 54.1 & -36.2 &        &          \\\hline  
    29-Jan 16:48 &  &  & 18 $\pm$ 3   & 0.30 $\pm$ 0.05   \\
    29-Jan 17:23 &  &  & 19 $\pm$ 3   & 0.35 $\pm$ 0.05   \\
    29-Jan 17:58 &  &  & 21 $\pm$ 4   & 0.25 $\pm$ 0.05   \\
    29-Jan 18:58 &  &  & 15 $\pm$ 3   & 0.45 $\pm$ 0.05   \\
    29-Jan 19:30 &  &  & 13 $\pm$ 2   & 0.55 $\pm$ 0.05   \\
    29-Jan 20:21 &  &  & 13 $\pm$ 2   & 0.67 $\pm$ 0.05   \\ \hline 
	\end{tabular}
\caption{Cloud Characteristics.\label{tbl2}}
	
\end{center}

\end{table*}

\clearpage

\begin{center}\textbf{Figure Captions}\end{center}

\noindent Figure \ref{fobs}. --- Images from VLT/SINFONI observations in bandpasses use to measure the surface (2.027--2.037\,$\mu$m), troposphere (2.123--2.128\,$\mu$m), and stratosphere (2.150--2.160\,$\mu$m), along with troposphere-probing images that have the stratospheric albedo subtracted (right column). The viewing geometry is shown with an outline of the bright continent, Xanadu (left column). Cloud features are indicated with colored arrows corresponding to the spectra in Figure \ref{fspec}.

\vspace{1cm}

\noindent Figure \ref{fspec}. --- VLT/SINFONI spectra centered on clouds. Raw spectra (dotted lines) as well as spectra that are corrected for the contribution from stratospheric aerosol (solid lines). (A) the optically thin stratospheric scattering (e.g., at $\lambda>2.16 \mu$m) has a larger contribution near the limb and complicates the direct comparison of cloud spectra in the 2.12--2.15\,$\mu$m region (dotted lines); this effect is removed (solid lines) by subtracting a spectrum  from the same spatial location in a cloud-free datacube, observed on 2007-02-23\,UT. (B) on the 28th and 30th the corrected cloud spectra are brighter near 2.15\,$\mu$m than on the 29th and 31st and systematically exceed the uncertainty (dashed line) beyond 2.16\,$\mu$m on these days, indicating that the cloud tops likely extend into the upper troposphere. Differences in 2.12\,$\mu$m albedo are indicative of variations in the cloud fill fraction and the relative contribution from clouds lower in the atmosphere. The corrected spectra are dominated by noise beyond 2.30\,$\mu$m. A histogram of $I/F$ from 2.30--2.45\,$\mu$m (C) is fit by a Gaussian and used to estimate the uncertainty in the spectrum.  Spectra are median sampled in a 5$\times$5 pixel window centered on the clouds.

\vspace{1cm}

\noindent Figure \ref{fCloudFit}. --- Images and spectra from calculations that best fit the altitudes and spatial extent of clouds in VLT/SINFONI observations on four nights. Images from 2.11--2.15\,$\mu$m are displayed on the leftmost panels to illustrate the difference between models with and without a cloud, along with a close-up of the locations of the maximum number of pixels used to model the cloud. The best fit spectra and observations are compared with the base (cloud-free) model in the second panel from the left. The figure of merit, $\Delta\widehat{\chi^{2}}$, used for comparing the calculations to the observations is displayed in the two right-hand columns for a range of optical depths and cloud altitudes. The 1$\sigma$ uncertainties in the parameters correspond to the $\Delta\widehat{\chi^{2}}$=1 contours.

\vspace{1cm}

\noindent Figure \ref{fVIMSobs}. --- Spacecraft observations from Cassini/VIMS of individual convective cells, e.g, at location marked (A) corresponding to spectra in panel (A), evolving over hourly timescales. Observation geometry is illustrated in the top row of images. Cloud features with stratospheric flux subtracted and rainbow color stretch for clarity indicated with observing time (middle). Spectra over three convective cells (A -- C), and a cloud-free location (D), shown in bottom panels. Labels for spectra (marked 1 through 6 ) correspond to images numbers (top). Spectra from locations A and C show steady increase in albedo, while in region B the change in flux is not monotonic with time. Changes in flux are due to both the variation in cloud-top altitude and the increasing optical depth of the clouds. Insets in panels (A) and (B) show an enlargement region near 2.175\,$\mu$m specified by the box around the spectra.

\vspace{1cm}

\noindent Figure \ref{fVIMSchi}. Constraints on cloud altitudes and opacities in the models used to interpret the observations. The figure of merit, $\Delta\widehat{\chi^{2}}$, over a range of altitudes from 1--60km, and optical depths from $\tau$=0.01 to 10. A narrower range of optical depths are plotted for clarity. Three columns correspond to the three cloud locations, labelled A, B, and C, with each row corresponding to a particular observation time, increasing from top to bottom. Dotted lines are shown at the same altitude and optical depth for each cloud location to be a guide for comparing the changes in cloud properties with time.

\vspace{1cm}

\noindent Figure \ref{fVIMSspecfit}. Comparison of Cassini/VIMS spectra (thick grey lines) with calculations of both clear-sky spectra (dashed lines) and calculations with clouds (thin black lines) that best fit the observations at 6 time-steps. Spectra are successively offset by 0.1 I/F for clarity.
Panels (A) through (C) show data for a particular cloud, identified by corresponding letters in Figure \ref{fVIMSobs}. Panel (D) illustrates the fit of clear-sky spectra to observations. The spatial locations of these spectra are plotted in panels (3) and (4) of Figure \ref{frepro}.
\vspace{1cm}

\noindent Figure \ref{frepro}. --- Cylindrical projection of cloud observations, panels (1) through (5). The observed I/F above an arbitrary threshold, indicating cloud location (dark grey smudges), is plotted for clarity. Observation time is indicated in the upper right of each panel. Crosses labelled (A) through (D) mark the locations of VIMS spectra used for the analysis of clouds, and correspond to the panels in Figures 4, 5, and 6. Location (D) is the clear-sky spectrum in the VIMS dataset. Xanadu is outlined (solid) and Hotei Arcus is indicated with the dashed curve along with a 350 km scale bar. The region of cloud formation is stationary, near 38$^{\circ}$S and 62$^{\circ}$W, for at least 34\,hrs; panels (1) through (3). On the final night of observations, 2007 Jan 31 UT, the region of cloud formation is shifted from the previous nights, and it is unclear how the regions are related.

\clearpage

\section*{Acknowledgments}

The authors wish to thank Mike Flasar, as well as two anonymous referees, for thoughtful comments on the manuscript. IdP and M\'A are supported by NSF and the Technology Center for Adaptive Optics, 
managed by the University of California at Santa Cruz under cooperative agreement 
AST-9876783 and NASA grant NNG05GH63G. Observations were performed at the VLT operated by the European Southern Observatory, under program ID 078.C-0458.

\clearpage

\clearpage

\clearpage

\clearpage

\clearpage

\clearpage

\clearpage

\end{document}